\documentclass[prd,twocolumn,superscriptaddress,nofootinbib,showpacs]{revtex4}

\usepackage{graphicx} 
\usepackage{dcolumn}  
\usepackage{amsmath}
\usepackage{bm}       
\usepackage{epsfig}
\usepackage{amsfonts}
\usepackage{amssymb}
\usepackage{revsymb}

\begin{document}

\title{A note on cosmological parameters and the topology of the universe}

\author{M.J. Rebou\c{c}as}\email{reboucas@cbpf.br} 
\affiliation{Centro Brasileiro de Pesquisas F\'{\i}sicas\\
Rua Dr.\ Xavier Sigaud 150 \\ 22290-180 Rio de Janeiro -- RJ, Brazil}

\author{J.S. Alcaniz}\email{alcaniz@on.br} 
\affiliation{Observat\'orio Nacional\\
Rua Gal.\ Jos\'e Cristino 77 \\
20921-400, Rio de Janeiro -- RJ, Brazil}

\date{\today}

\begin{abstract}
Given the wealth of increasingly accurate cosmological observations, 
especially the recent results from the WMAP, and the development 
of methods and strategies in the search for cosmic topology, it is 
reasonable to expect that we should be able to detect the spatial 
topology of the Universe in the near future. Motivated by this, we 
examine to what extent a possible detection of a nontrivial topology 
of positively curved universe may be used to place constraints on 
the matter content of the Universe. We show through concrete 
examples that the knowledge of the spatial topology allows to 
place constraints on the density parameters associated to dark matter 
($\Omega_m$) and dark energy  ($\Omega_{\Lambda}$).
\end{abstract}


\maketitle

\section{Introduction}
\label{Intro} 
Questions such as whether the Universe will expand forever or  
eventually re-collapse and end with a Big Crunch, and what its 
shape and size may be, are among the most fundamental challenges 
in cosmology. Regarding the former question, it is well known 
that the ultimate fate of the Universe is intrinsically associated 
with the nature of its dominant components. In the 
Friedmann-Lema\^{\i}tre-Robertson-Walker (FLRW) class of models, 
for instance, a universe that is dominated by a pressureless fluid 
(as, e.g., baryons and/or dark matter) or any kind of fluid with 
positive pressure (as radiation, for example) will expand forever 
if its spatial geometry is Euclidean or  hyperbolic, or will 
eventually re-collapse if it is spherical. 
This predictable destiny for the Universe, however, 
may be completely modified if it is currently dominated by some sort 
of negative-pressure dark component, as indicated by a number of 
independent observational results (see, e.g., Ref.~\cite{revde}). 
In this case, not only the dynamic but also the thermodynamic fate of 
the Universe may be completely different, with the possibility of an 
eternally expanding closed model~\cite{closed}, an increasingly hot 
expanding universe~\cite{limajas} or even a progressive rip-off of the 
large and small scale structure of matter ending with the occurrence 
of a curvature singularity, the so-called big smash~\cite{caldwell}.

The remaining questions, concerning the shape and size of our 
$3$--dimensional world, go in turn beyond the scope of general 
relativity (GR), since they have an intrinsically topological 
nature. In this way, approaches or answers to these questions 
are ultimately associated with measurements of the \emph{global} 
structure (topology) of the Universe and, as a \emph{local} metric 
theory, GR cannot say much about it, leaving the global topology of 
the Universe undetermined%
\footnote{In this paper, in line with the usage in the literature,
by topology of the universe we mean the topology of its spatial 
section.}.
Over the past few years, several aspects of the cosmic topology have 
become topical (see, e.g., the review articles Ref.~\cite{CosmTopReviews}), 
given the wealth of increasingly accurate cosmological observations, especially 
the recent results from the Wilkinson Microwave Anisotropy Probe (WMAP) 
experiment~\cite{WMAP}, which have heightened the interest in the 
possibility of a universe with a nontrivial spatial topology%
\footnote{For instance, the reported non-Gaussianity in cosmic 
microwave background (CMB) maps~\cite{CMB+NonGauss} along with the 
small power of large-angle fluctuations~\cite{WMAP-Spergel}, and some 
features in the power spectrum~\cite{CMB+NonGauss,WMAP-Spergel,Align} 
are all large-scale anomalies which have been suggested as potential
indications of a universe with nontrivial topology~\cite{Poincare&More,Aurich12}.
See Ref.~\cite{Copi-etal05} for a detailed discussion on these
anomalies.}.

A pertinent question the reader may ask at this point is whether 
the current values of cosmological density parameters, which help 
us to answer the above first question (associated with the ultimate 
fate of the Universe), can be constrained by a possible detection 
of the spatial topology of the Universe. Our primary objective here 
is to address this question by focusing our attention on possible 
topological constraints on the  density parameters associated with 
the baryonic/dark matter  ($\Omega_m$) and dark energy ($\Omega_{\Lambda}$). 
Motivated by the best fit value  for the total energy density 
$\Omega_{\rm{T}} \equiv  \Omega_m + \Omega_{\Lambda} = 1.02 \pm\, 0.02$ 
($1\sigma$ level) reported by WMAP team \cite{WMAP-Spergel}, which 
includes a positively curved universe as a realistic possibility, we 
shall consider globally  homogeneous spherical manifolds, some of which 
account for the suppression of power at large scales observed by 
WMAP~\cite{WMAP-Spergel}, and also fits the WMAP temperature two-point 
correlation function~\cite{Aurich12,Aurich3}. To this end, in the next 
section we present our basic context and prerequisites, while in the 
last section we discuss our main results and present some concluding 
remarks. 

\section{Basic Context and Prerequisites} 

Within the framework of standard cosmology, the universe is described 
by a space-time manifold $\mathcal{M}_4 = \mathbb{R} \times M$  with a 
locally homogeneous and isotropic Robertson--Walker (RW)  metric
\begin{equation}
\label{RWmetric} ds^2 = -dt^2 + a^2 (t) \left [ d \chi^2 + f^2(\chi)
(d\theta^2 + \sin^2 \theta \, d\phi^2) \right ] \;,
\end{equation}
where $f(\chi)=(\chi\,$, $\sin\chi$, or $\sinh\chi)$ depends on the 
sign of the constant spatial curvature ($k=0,1,-1$, respectively). 
The $3$--space $M$ is usually taken to be one of the following 
simply-connected spaces: Euclidean $\mathbb{R}^3$, spherical 
$\mathbb{S}^3$, or  hyperbolic $\mathbb{H}^3$. However, given that 
the simple-connectedness of our space $M$ has not been established, 
our $3$--space may equally well be any one of the possible quotient 
manifolds $M = \widetilde{M}/\Gamma$, where $\Gamma$ is a fixed 
point-free group of isometries of the covering space $\widetilde{M}=(\mathbb{R}^{3},\mathbb{S}^{3}, \mathbb{H}^{3})$. 
Thus, for example, in a universe whose geometry of the spatial 
section is Euclidean ($k=0$), besides $\mathbb{E}^{3}$ there are 
6 classes of topologically distinct compact orientable $3$--spaces 
$M$ that admits this geometry, while for universes with either 
spherical ($k=1$) and hyperbolic ($k=-1$) spatial geometries there 
is an infinite number of topologically non-homeomorphic (inequivalent) 
manifolds with nontrivial topology that can be endowed with these 
geometries.  
 
Quotient manifolds are compact in three independent directions, 
or compact in two or at least one independent direction. In 
compact manifolds, any two given points may be joined by more 
than one geodesic. 
Since the radiation emitted by cosmic sources follows geodesics, 
the immediate observational consequence of a nontrivial detectable 
spatial topology of $M$ is that the sky may show multiple images of 
radiating sources: cosmic objects or specific correlated spots of 
the cosmic microwave background radiation (CMBR). At very large 
scales, the existence of these multiple images (or pattern repetitions) 
is a physical effect that can be used to probe the $3$-space topology. 
In this work, we use the so-called ``circles-in-the-sky"  method
(for cosmic crystallographic methods see, e.g., Refs.~\cite{CCmethods}), 
which relies on multiple copies of correlated circles in the CMBR
maps~\cite{CSS1998}, whose existence is clear from the following
reasoning: 
In a space with a detectable nontrivial topology, the last 
scattering sphere (LSS) intersects some of its topological images 
along pairs of circles of equal radii, centered at different points 
on the LSS, with the same distribution 
of temperature fluctuations, $\delta T$. 
Since the mapping from the LSS to the night sky sphere preserves 
circles~\cite{CGMR05}, these pairs of matching circles will be 
inprinted on the CMBR temperature fluctuations sky maps regardless 
of the background geometry and detectable topology.
As a consequence, to observationally probe a nontrivial topology 
on the available largest scale, one should scrutinize the full-sky CMB
maps in order to extract the correlated circles, whose angular
radii and relative position of their centers can be used to
determine the topology of the universe. Thus, a nontrivial
topology of the space section of the universe may be observed,
and can be probed through the circles-in-the-sky for all locally
homogeneous and isotropic universes with no assumption on the
cosmological density parameters.

Let us now state our basic cosmological assumptions and fix some
notation. In addition to the RW metric (\ref{RWmetric}), we assume 
that the current matter content of the Universe is well approximated 
by Cold Dark Matter (CDM) of density $\rho_m$ plus a cosmological 
constant $\Lambda$.
In this standard $\Lambda$CDM context,  for nonflat spaces the 
scale factor $a(t)$ can be identified with the curvature radius 
of the spatial section of the universe at time $t=t_0$, which
is given by
\begin{equation}
\label{CurvRad}
a_0^2 = \frac{k}{H_0^2(\Omega_{\rm{T}}-1)} \; ,
\end{equation}
where here and in what follows the subscript $0$ denotes evaluation 
at present time $t_0$, $H_0$ is the Hubble constant, and $\Omega_{\rm{T}}$ 
is the total density at $t=t_0$. 
In this way, for nonflat spaces the distance $\chi$ of any point 
with coordinates $(\chi, \theta, \phi)$ to the origin (in the 
covering space) \emph{in units of the curvature radius}, $a_0=a(t_0)$,
reduces to   
\begin{equation}
\label{redshift-dist}
\chi = \sqrt{|\Omega_k|} \int_1^{1+z} \hspace{-4mm}
\frac{dx}{\sqrt{x^3 \Omega_{m0} + x^2 \Omega_k + 
 \Omega_{\Lambda 0}}} \; ,
\end{equation}
where $x=1+z$ is an integration variable, and $\Omega_k = 1-\Omega_{\rm{T}} $. 
Throughout this paper we shall measure the lengths in unit of
curvature radius $a_0$.

A typical characteristic length of nonflat manifolds $M$, which we 
shall use in this paper, is the so-called injectivity radius $r_{inj}$, 
which is defined as  the radius of the smallest sphere `inscribable' 
in $M$. An important mathematical result is that $r_{inj}$, expressed 
in terms of the curvature radius, is a constant (topological invariant) 
for any given spherical and hyperbolic manifolds. 

In this work we shall focus our attention in globally homogeneous 
spherical manifolds, as presented  in Table~\ref{SingleAction} (see 
also its caption  for more details). These manifolds satisfy a 
topological principle of homogeneity, in the sense that all points 
in $M$ are topologically equivalent.

\begin{table}[!hbt]
\begin{center}
\begin{tabular}{|cccc|}
\hline
\ Name  & Covering Group $\Gamma$ & \ Order of $\Gamma$ \ & \ $r_{inj}$ \  \\
\hline \hline
$\mathcal{Z}_n$ &   Cyclic              $Z_n$   & $n$  & $\pi/n$           \\
$\mathcal{D}_m$ &   Binary dihedral     $D^*_m$ & $4m$ & $\pi / 2m $      \\
$\mathcal{T}$   & \ \ Binary tetrahedral  $T^*$    & 24   & $ \pi/6$          \\
$\mathcal{O}$   & \ \ Binary octahedral   $O^*$    & 48   & $\pi/8$          \\
$\mathcal{D}$   & \ \ Binary icosahedral  $I^*$    & 120  & $\pi/10$         \\
\hline
\end{tabular}
\end{center}
\caption{The globally homogeneous spherical manifolds are
of the form $M=\mathbb{S}^3/\Gamma$. The first column gives the 
name we use for the manifolds. The second column displays the covering 
groups $\Gamma$. Finally, the remaining columns present the 
order of the group $\Gamma$ and the injectivity radius $r_{inj}$. 
The cyclic and binary dihedral cases actually constitute families of 
manifolds, whose members are given by the different values of the 
integers $n$ and $m$. The order of $\Gamma$ gives the number of
fundamental polyhedra  needed to fulfill the whole covering space 
$\mathbb{S}^3$. Thus, for example, for the manifold $\mathcal{D}$
which is the the well-known Poincar\'e dodecahedral space, the 
fundamental polyhedron is a regular spherical dodecahedron, 
$120$ of which tile the $3$--sphere into identical cells that  
are copies of the FP.}
\label{SingleAction}
\end{table}

\section{Main Results and Final Remarks}
\label{Finals}
\begin{figure}[t]
\centerline{\psfig{figure=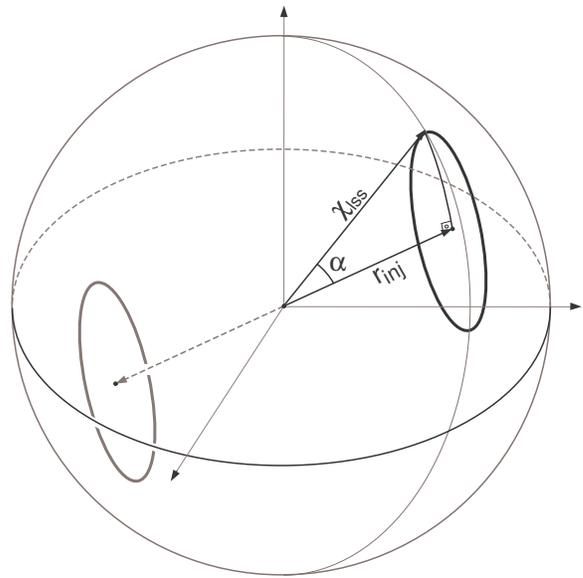,width=3.2truein,height=3.2truein,angle=0}
\hskip 0.1in} \caption{A schematic illustration of two antipodal
matching circles in the sphere of last scattering. These pair of circles 
occurs in all globally homogeneous positively curved manifolds with a 
detectable nontrivial topology. The relation between the angular radius 
$\alpha$ and the angular sides $r_{inj}$ and $\chi^{}_{lss}$ is given 
$\cos \alpha = \tan r_{inj}\, \cot \chi^{}_{lss}$.}
\label{CinTheSky1}
\end{figure}

To investigate the extent to which a possible detection of a nontrival
topology may place constraints on the cosmological density parameters, 
we consider here the globally homogeneous spherical manifolds. In these 
$3$--spaces the number of pairs of matching circles depends on the 
ratio of the injectivity radius $r_{inj}$ to the radius $\chi^{}_{lss}$ of 
LSS, which in turn depends on the density parameters (see Ref.~\cite{Aurich3} 
for examples of specific estimates of this number regarding $\mathcal{T}$, 
$\mathcal{O}$ and $\mathcal{D}$). Nevertheless, if the topology of a 
globally homogeneous spherical manifold is detectable%
\footnote{The extent to which a nontrivial topology may or may not
be detected has been discussed in references~\cite{TopDetec}.}
the correlated pairs will be antipodal, i.e. the centers of correlated 
circles are separated by $180^\circ$, as shown in Figure~\ref{CinTheSky1}.

Clearly the distance between the centers of each pair of the {\it first}
correlated circles is twice the injectivity radius $r_{inj}$.
Now, a straightforward use of known  trigonometric rules to the right-angled 
spherical triangle shown in Figure~\ref{CinTheSky1} yields a relation
between the angular radius $\alpha$ and the angular sides $r_{inj}$ and
radius $\chi^{}_{lss}$ of the last scattering sphere, namely
\begin{equation}
\label{cosalpha}
\cos \alpha = \frac{\tan r_{inj}}{\tan \chi^{}_{lss} }\;,
\end{equation}
where $r_{inj}$ is a topological invariant, whose values are given
in Table~\ref{SingleAction}, and the distance $\chi^{}_{lss}$ of
the last scattering surface to the origin in units of the curvature
radius is given by~(\ref{redshift-dist}) with $z_{lss}=1089$%
~\cite{WMAP-Spergel}.

Equations~(\ref{cosalpha}) along with (\ref{redshift-dist}) give 
the relations between the angular radius $\alpha$ and the cosmological 
density parameters $\Omega_{\Lambda}$ and $\Omega_{m}$, and thus can be
used to set bounds on these parameters. To quantify this we
proceed in the following way. Firstly, as an example, we assume the 
angular radius $\alpha = 50^\circ$.
Secondly, since the measurements of the radius $\alpha$
unavoidably involve observational uncertainties, in order
to obtain very conservative results we take $\delta {\alpha} \simeq
6^\circ$.%
\footnote{These values are taken as typical and for the sake of
comparison with the results of Refs.~\cite{Previous}. We note,
however, that our general conclusions hold regardless of the
precise value for $\alpha$ and its uncertainty.}

In order to study the effect of the cosmic topology on the density
parameters $\Omega_{m}\,$ and $\,\,\Omega_{\Lambda}\,$, we consider 
the binary tetrahedral $\mathcal{T}$ and the binary octahedral $\mathcal{O}$ 
spatial topologies (see Table~\ref{SingleAction}), to reanalyze with
these two topological priors the constraints on these parameters that arise 
from the so-called \emph{gold} sample of 157 SNe Ia, as compiled by 
Riess \emph{et al.\/}~\cite{Riess2004}, along with the latest Chandra 
measurements of the X-ray gas mass fraction in 26 X-ray luminous, 
dynamically relaxed galaxy clusters (spanning the redshift range 
$0.07 < z < 0.9$) as provided by Allen {\it et al.}~\cite{Allen2004} 
(see also \cite{alcxray} for details on SNe Ia and X-ray statistics).
The  $\mathcal{T}$ and $\mathcal{O}$ spatial topology is added 
to the conventional SNe Ia plus clusters data analysis as a Gaussian 
prior on the value of $\chi^{}_{lss}$, which can be easily obtained 
from an elementary combination of~(\ref{cosalpha}) and~(\ref{redshift-dist}). 
In other words, the contribution of the topology to $\chi^2$ is a term 
of the form $\chi^2_{\mathrm{topology}} = (\chi^{\mathrm{Obs}}_{lss} - 
\chi^{\mathrm{Th}}_{lss})^2 / (\delta \chi_{lss})^2$.

\begin{figure*}[t]
\centerline{
\psfig{figure=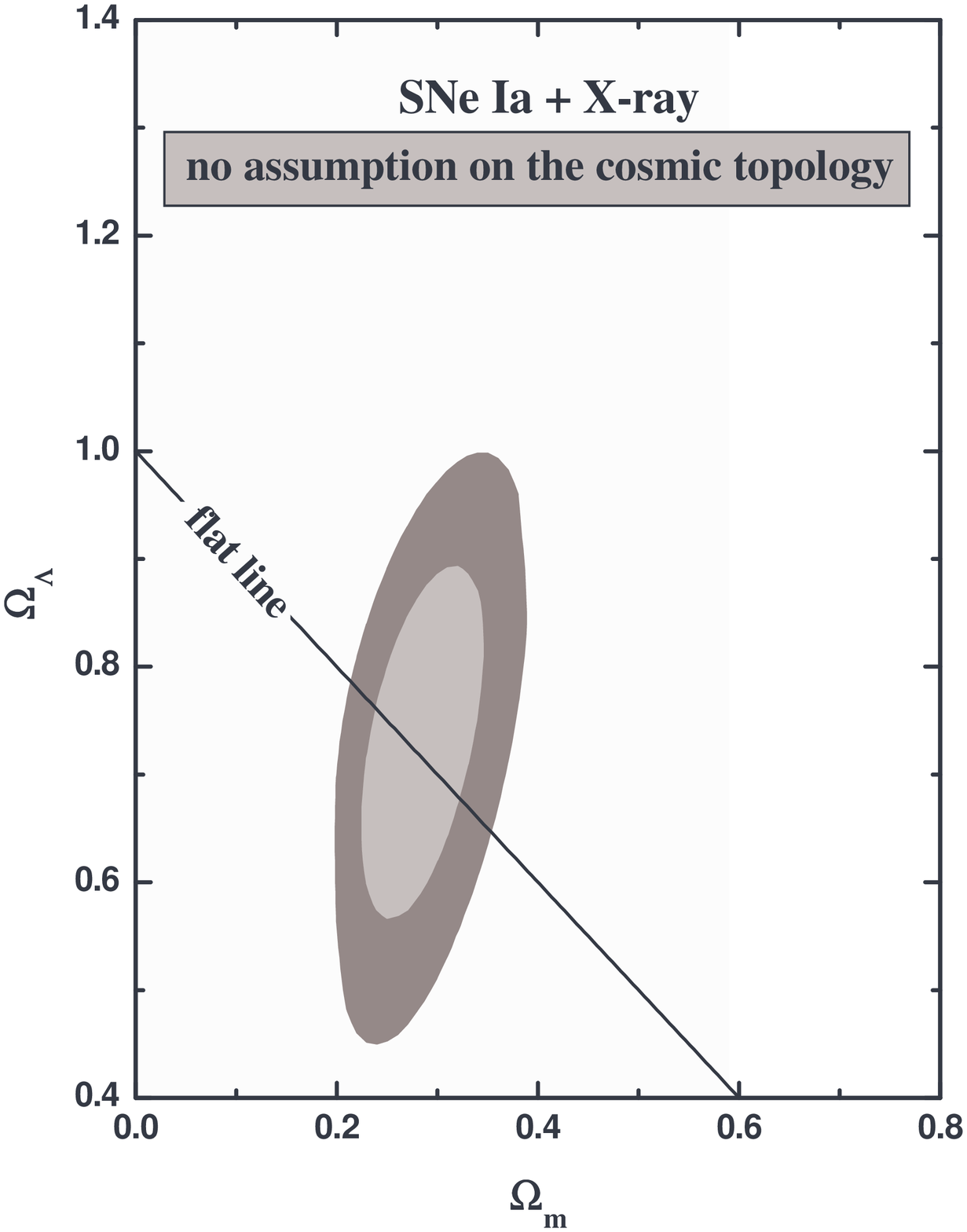,width=2.4truein,height=3.5truein,angle=0} \hskip -0.2cm
\psfig{figure=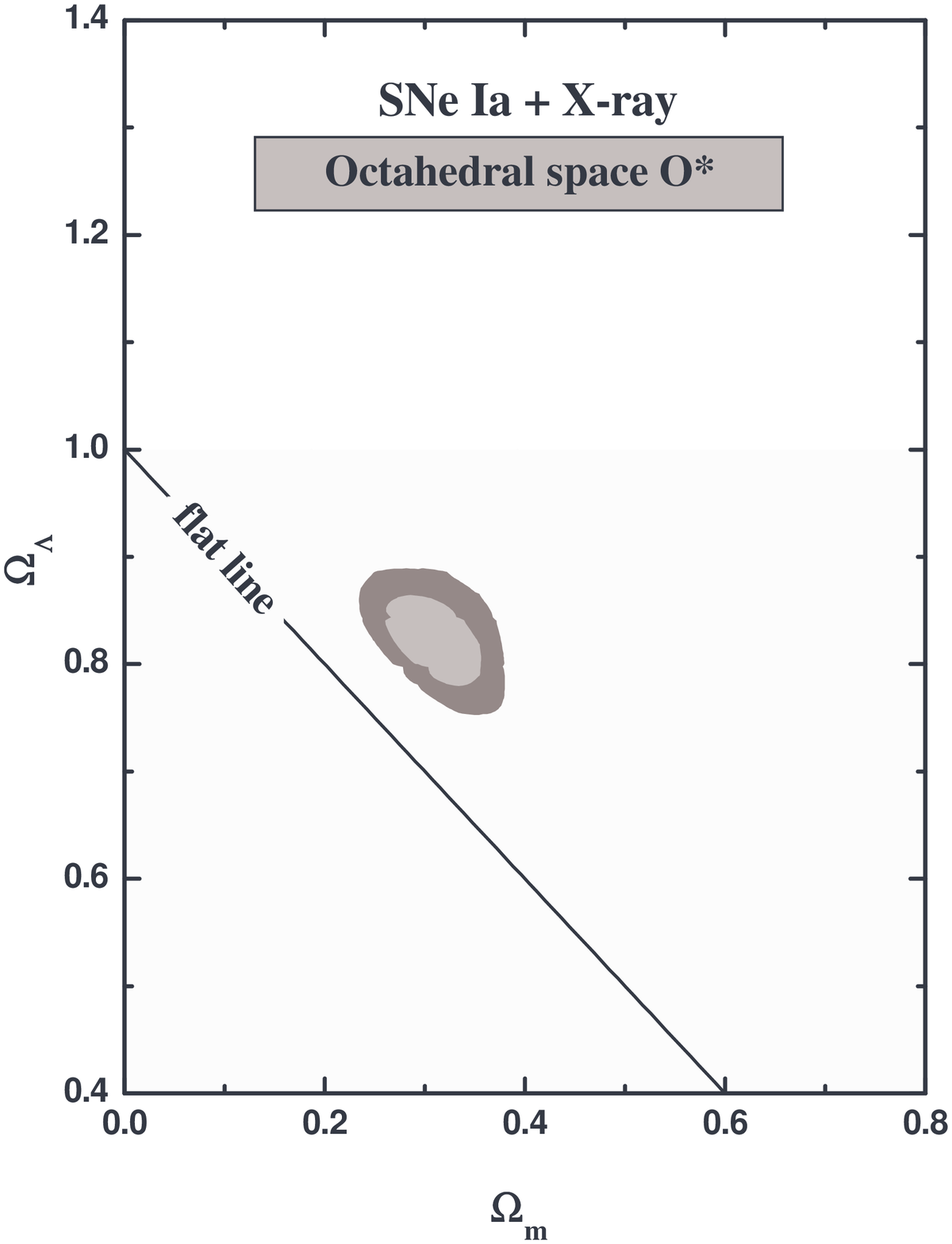,width=2.4truein,height=3.5truein,angle=0} \hskip -0.2cm
\psfig{figure=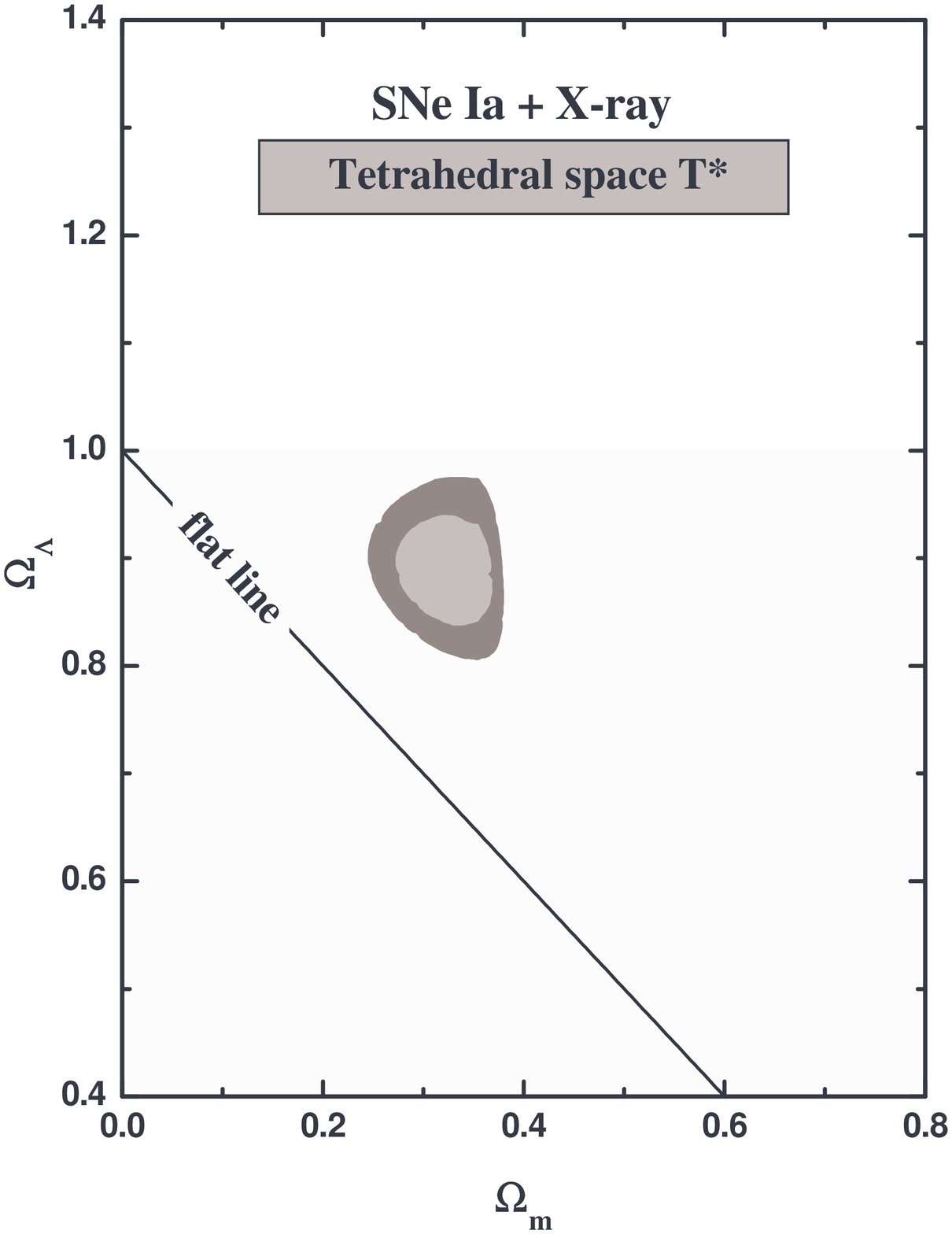,width=2.4truein,height=3.5truein,angle=0}\vspace{-1cm}}
\caption{The results of our statistical analyses. The panels show confidence 
regions ($68.3\%$ and $95.4\%$ c.l.) in the $\Omega_m - \,\,\Omega_{\Lambda}$ 
plane from the latest Chandra measurements of the X-ray gas mass fraction in 
26 galaxy clusters ($0.07 < z < 0.9$) plus determinations of the baryon density 
parameter, measurements of the Hubble parameter and the \emph{gold} sample of 
157 SNe Ia. {\bf{Left Panel:}} The conventional SNe Ia plus X-ray analysis 
shown here for the sake of comparison.
{\bf{Central Panel:}} A binary octahedral space topology is assumed as the 
global topology of the Universe. The best fit values for the matter and vacuum 
density parameters are, respectively, $\Omega_m = 0.31 \pm 0.05$ and 
$\Omega_{\Lambda} = 0.82 \pm 0.05$ ($\Omega_{\mathrm{T}} 
= 1.13 \pm 0.07$) at 95.4\% c.l..  
{\bf{Right Panel:}} A binary tetrahedral spatial topology is assumed. 
At 95.4\% c.l., the best fit values for the matter and vacuum density 
parameters are $\Omega_m = 0.32 \pm 0.06$, $\,\Omega_{\Lambda} = 0.89 \pm 0.06$,
and $\Omega_{\mathrm{T}} = 1.21 \pm 0.08$.} 
\label{statisticalAnalysis}
\end{figure*}

Figures 2b and 2c (central and right panels) show the results of our 
statistical analysis. Confidence regions -- 68.3\% and 95.4\% confidence 
limits (c.l.) -- in the parametric space $\Omega_{m}$--$\,\,\Omega_{\Lambda}\,$ 
are displayed for the above described combination of observational data. 
For the sake of comparison, we also show in Fig.~2a the 
$\Omega_{m}$--$\,\,\Omega_{\Lambda}\,$ plane for 
the conventional SNe Ia plus Galaxy Clusters analysis, i.e., the one without 
the above cosmic topology assumption. By comparing both analyses, it is clear 
that a nontrivial space topology reduces considerably the parametric space 
region allowed by the current observational data, and also breaks some 
degeneracies arising from the current SNe Ia and X-ray gas mass fraction 
measurements. At 95.4\% c.l. our SNe Ia+X-ray+Topology analysis provides 
$\Omega_m = 0.31 \pm 0.05$ and $\Omega_{\Lambda} = 0.82 \pm 0.05$ (binary 
octahedral $O^*$) and $\Omega_m = 0.32 \pm 0.06$ and 
$\Omega_{\Lambda} = 0.89 \pm 0.06$ (binary tetrahedral  $T^*$).

Concerning the above analysis it is worth emphasizing  three important aspects. 
First, that the best-fit values depend weakly on the value used for radius 
$\alpha$ of the circle. Second, the uncertainty $\delta \alpha$ alters 
predominantly the area corresponding to the confidence regions, without 
having a significant effect on the best-fit values. Third, we also notice that 
there is a topological degeneracy in that the same best fits and 
confidence regions for, e.g., the $\mathcal{T}$ topology, would equally 
arise from either $\mathcal{Z}_6$ or $\mathcal{D}_3$ spatial topology. 
Similarly, $\mathcal{O}$,  $\mathcal{Z}_8$ and $\mathcal{D}_3$ give 
rise to identical bounds on the density parameters. This kind of 
topological degeneracy passed unnoticed in Refs.~\cite{Previous}. 

Finally, we emphasize that given the wealth of increasingly accurate 
cosmological observations, especially the recent results from the WMAP, 
and the development of methods and strategies in the search for cosmic 
topology, it is reasonable to expect that we should be able to detect it.
Besides it importance as a major scientific achievement, we have shown 
through concrete examples that the knowledge of the spatial topology 
allows to place constraints on the density parameters associated to 
dark matter ($\Omega_m$) and dark energy  ($\Omega_{\Lambda}$).

\section*{Acknowledgements}
We thank CNPq for the grants under which this work was carried out. 
We also thank A.F.F. Teixeira for the reading of the manuscript and 
indication of relevant misprints and omissions.

\end{document}